\documentclass[12pt]{iopart}
\usepackage{graphics,epsfig}
\begin{document}

\title{A c-theorem for the entanglement entropy}

\author{H Casini and M Huerta}

\address{Centro At\'omico Bariloche,
8400-S.C. de Bariloche, R\'{\i}o Negro, Argentina}
\eads{\mailto{casini@cab.cnea.gov.ar},  
\mailto{huerta@cabtep2.cnea.gov.ar}}
\begin{abstract}
 The combination of the Lorentz symmetry and the strong
subadditive property of the entropy leads to a c-theorem for the
entanglement entropy in 1+1 dimensions. We present a simple derivation of this theorem 
and compare the associated
c-functions with the Zamolodchikov's
ones for the case of free fields. We discuss the various difficulties which obstacle the naive
generalizations of the entropic c-theorem to higher dimensions.
\end{abstract}


\section{Entanglement entropy}

Suppose we are interested in the physics of events localized in a region $A$ of the space. The Hilbert space of states can be decomposed accordingly as a tensor product $
 {\cal H}={\cal H}_{A}\otimes {\cal H}_{-A}$ where ${\cal H}_{A}$ and  ${\cal H}_{-A}$ are the spaces of states localized in $A$ and in the complementary region $-A$ respectively. Take the vacuum (the fundamental state) as a global state  $\left| \Psi \right>$ of the system with density matrix $
\rho_0= \left| \Psi \right> \left< \Psi \right|$. 
 The state $\rho_A$ relevant to the algebra of operators acting on ${\cal H}_{A}$
  follows from the partial trace of $\rho_0$ over the Hilbert space ${\cal H}_{-A}$. That is, we obtain  the local reduced density matrix 
\begin{equation}
 \rho_{A}=\textrm{tr}_{{\cal H}_{-A}}\left| \Psi \right> \left< \Psi \right|\,.
\end{equation}
The global state $\rho_0$ is generally entangled in the bipartite system ${\cal H}_{A}\otimes {\cal H}_{-A}$ and in consequence this matrix is a mixed (non pure) state for the local region.  
The corresponding entropy
 \begin{equation}
 S(A)=-\textrm{tr}(\rho_A\log \rho_A)\,
\end{equation}
is usually called entanglement or geometric entropy. 
  
The entanglement entropy is one of the most prominent candidates to explain the intriguing entropy of the black holes \cite{bombelli}. However, in this proposal the role of quantum gravity is fundamental to produce a finite entropy, and the whole subject is still controversial.  In a series of recent papers a definite conjecture for the meaning of the entanglement entropy in the context of the AdS-CFT duality has been given \cite{ryu}.

This and other measures of entanglement have also been extensively studied in condensed matter and low dimensional systems, partially motivated by advances in quantum information theory and the density matrix renormalization group method.  As a result it was uncovered that a variety of phenomena have an interesting  correlation with the entanglement properties of fundamental states. This includes quantum phase transitions \cite{phase}, topological order \cite{topological}, and, as will be reviewed in this paper, the renormalization group irreversibility in two dimensions \cite{ch1,rg}. 

\section{Entanglement entropy and quantum field theory}
From the point of view of quantum field theory (QFT) the entanglement entropy $S(A)$ can be considered as a non local variable which is definable for any theory disregarding the field content. In this sense it can be identified as a natural tool to investigate general properties of QFT. In fact, the entanglement entropy satisfies very remarkable non perturbative relations (see for example \cite{h1})  
\begin{eqnarray}              
 S(A)=S(-A)\,,\\
S(A)+S(B)\ge S(A\cup B) +S(A\cap B)\,.
\end{eqnarray}
The first one follows from the purity of the vacuum state, while the second, called strong subadditivity (SSA), is a general and very important property of the entropy for multipartite systems. 

From a technical point of view, the entanglement entropy $S(A)$ is proportional to the variation of the Euclidean free energy with respect to the introduction of a small  conical singularity at the boundary of $A$ \cite{conical,conformal}. Explicitly we have 
\begin{equation}
S(A)=\lim_{\alpha\rightarrow 1} \left(1-\frac{d}{d\alpha}\right) \log Z(A,\alpha),\label{alpha}
\end{equation}
where $Z(A,\alpha)$ is the Euclidean partition function of the theory on a space with conical singularity of angle $2\pi \alpha$  at the boundary of $A$. 

The entanglement entropy in the continuum theory is ultraviolet divergent. However, it has a nice geometrical structure of divergences. In $d$ spatial dimensions we have  \cite{ch2}
 \begin{equation}
 S(V)=g_{d-1}[\partial V]\,\epsilon^{-(d-1)}+...+ g_1[\partial V]\,\epsilon^{-1} + g_0[\partial V]\,\log (\epsilon)+ S_0(V)\,,   \label{div}
 \end{equation}
 where $S_0(V)$ is a finite part, $\epsilon$ is a short distance cutoff, and the $g_i$ are local and extensive  functions on the boundary $\partial V$, which are homogeneous of degree $i$. The leading divergent term coefficient  $g_{d-1}[\partial V]$ is then proportional to the $(d-1)$ power of the size of $V$, and this is usually referred to as the area law for the entanglement entropy \cite{srednicki}. However, $g_i$ for $i> 0$ depends on the regularization procedure and $g_{d-1}$ is not proportional to the area if this later is not rotational invariant. These terms are not physical within QFT since they are not related to continuum quantities.

Universal quantities are present in $S(A)$ however. In particular the mutual information 
\begin{equation}
I(A,B)=S(A)+S(B)-S(A\cup B)
\end{equation}
between two non-intersecting regions is universal since the boundary terms get subtracted away \cite{ch1}. In fact, a general theorem based on the strong subadditive property and the Lorentz covariance \cite{h1} establishes that if the entanglement entropy is finite, the mutual information is identically zero for a relativistic QFT. The presence of a non zero mutual information implies that the entropy cannot be made finite through contributions coming from  the physics at the high energy sectors (within QFT).
 
\section{The c-theorem}
The statement of the c-theorem in $1+1$ dimensions \cite{zamo} can be expressed in two equivalent ways:

\bigskip

\noindent (a) There is a universal dimensionless function of the couplings $c(\{\lambda_i \})$ in the theories space in $1+1$ dimensions which is non-increasing along the renormalization group trajectories and stationary at the fixed points, where it takes a finite value proportional to the Virasoro central charge $C_V$.

\bigskip

\noindent (b) For any theory in $1+1$ dimensions there is a universal dimensionless function of a distance $c(r)$ which is non-increasing under dilatations and takes a finite value proportional to $C_V$ at the fixed points.

\bigskip 

An interesting interpretation which follows from the formulation (a) is that the c-function measures some kind of entropy related to the information which is lost during the renormalization group transformations.  However, if we take this too literally, we could conclude that the theorem should also be valid outside the domain of unitary and relativistic theories, where it actually does not hold. In contrast, the formulation (b) does not refer to the renormalization group and show the c-theorem as a property of the continuum theory, disregarding the procedure used to define it.

The Zamolodchikov's proof of the c-theorem is given in the Euclidean formulation of the two dimensional QFT, and involves the covariance of the energy momentum tensor correlators plus the reflexion positivity property. Here we show a very simple alternative proof based on the strong subadditive property of the entanglement entropy and the Lorentz invariance in a real space-time formulation \cite{ch1}.

 Consider two intervals of lengths $b$ and $c$ relatively boosted to each other and located as shown in figure 1. The causal domain of dependence of these intervals (causal shadow) have intersection and causal union given by the ones corresponding to the two intervals of length $a$ and $d$. These sets appear in the small side of the relativistic version of the strong subadditive inequality between $b$ and $c$ \cite{h1}. Then, we have\footnote{To be explicit, this relation is just the standard SSA inequality on the (non time-like) surface $c_1\cup a \cup b_1$ (see figure 1), which gives $S(c_1\cup a) +S(a\cup b_1)\ge S(a)+S(c_1\cup a\cup b_1)$, followed by the identification of $S(c_1\cup a)=S(c)$, $S(a\cup b_1)=S(b)$, and $S(c_1\cup a\cup b_1)=S(d)$, which is a consequence of the unitarity of the causal evolution.}  
\begin{equation}
S(c)+S(b)\ge S(a)+S(d)\,.\label{sss}
\end{equation}
The relativistic geometry gives the simple relation 
\begin{equation}
a\,d=c\,b\,,
\end{equation}
which can be rewritten as
\begin{equation}
c=\lambda a,\hspace{1cm} d=\lambda b,\hspace{1cm}\lambda=\frac{c}{a}=\frac{d}{b}\geq 1 \,.
\end{equation}
Therefore  (\ref{sss}) gives
\begin{equation}
S(b)-S(a)\ge S(\lambda b)-S(\lambda a)\,.\label{catorce}
\end{equation}
This means that the difference of the entropies $S(b)-S(a)$, with $b\ge a$ is non-increasing under dilatations. It is also dimensionless and universal since according to (\ref{div}) in two dimensions the divergent term for an interval is a constant (independent of the interval size) proportional to $\log (\epsilon )$.  
At the conformal fixed points the entropy can be explicitly evaluated \cite{conformal}, and is given by 
\begin{equation}
S(r)=\frac{C_V}{3} \log (r/\epsilon)\,.
\end{equation}
Thus, at the fixed points $S(b)-S(a)=(C_V/3) \log(b/a)\ge 0$. We conclude that  $S(b)-S(a)$ is a c-function for any $b > a$.

\begin{figure}[t]
\centering
\leavevmode
\epsfysize=5cm
\epsfbox{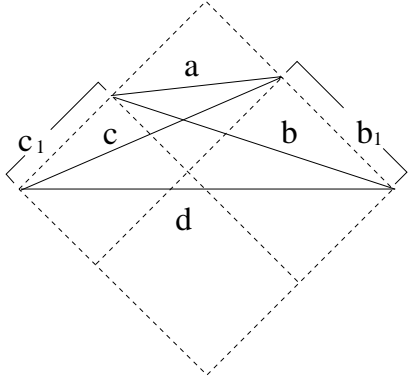}
\caption{Time is the vertical axis, the horizontal axis is the spatial coordinate $x$ and the null lines are drawn at $45^\circ$. The causal  domain of dependence (diamond shaped set drawn with dashed lines) corresponding to the spatial intervals $b$, $c$  have intersection given by the domain of dependence of $a$ and union (followed by causal completion) given by the domain of dependence of $d$.}
\end{figure} 
 
An entropic c-function depending on a single parameter (and containing all the information in $S(b)-S(a)$) can be defined by 
\begin{equation}
c(r)=r\frac{dS(r)}{dr}, 
\end{equation}
which is dimensionless, universal, positive, and according to (\ref{catorce}) satisfies 
\begin{equation}
c^{\prime}(r)=r\,S^{\prime\prime}(r)+S^{\prime}(r)\le 0 \,.
\end{equation}

\begin{figure}[t]
\centering
\leavevmode
\epsfysize=5cm
\epsfbox{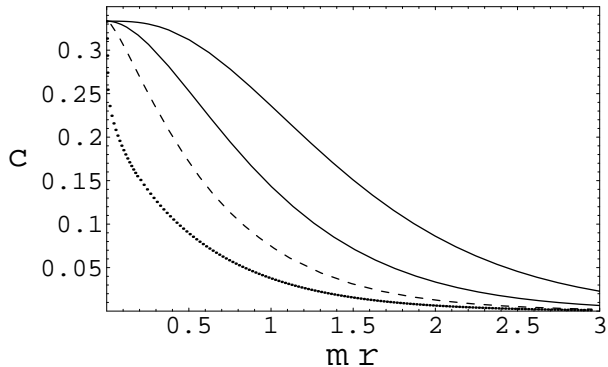}
\caption{From top to bottom: one third of the Zamolodchikov c-functions for a real scalar and a Dirac field, and entropic c-functions for a Dirac (dashed curve) and a real scalar field (dotted curve).}
\end{figure}

We have computed the entropic $c$-functions for free fermions and bosons numerically in a lattice and found analytically short and long distance expansions using (\ref{alpha}) \cite{ch3}. The analytic results involve sine-Gordon model correlators of exponential operators which are given in terms of Painlev\'e V differential equations. As shown in figure 2 these are quite different from the Zamolodchikov's ones. The leading short and long distance terms read
\begin{eqnarray}
c_D(t)\sim \frac{1}{3} -\frac{1}{3} t^2 \log^2 (t)\,\,\,\,\, \textrm{for}\,\, t\ll 1\,\,; \hspace{.4cm} c_D(t)\sim \frac{1}{2} t K_1(2 t)\,\,\,\,\, \textrm{for}\,\,t\gg 1\,, \\
c_S(t)\sim \frac{1}{3}+\frac{1}{2 \log (t)}\,\,\,\,\, \textrm{for}\,\,t\ll1\,\,; \hspace{.4cm} c_S(t) \sim \frac{1}{4}  t K_1(2 t)\,\,\,\,\, \textrm{for}\,\,t\gg 1\,,
\end{eqnarray}
where $t=m r$, $m$ is the field mass, and  $c_D(t)$ and $c_S(t)$ are the entropic c-functions corresponding to a free Dirac and a real scalar field respectively.  

\section{Is there an entropic c-theorem in more dimensions?}
Though several authors have made much progress in the direction of extending the c-theorem to higher dimensions using correlators of the energy momentum tensor, no definitive result has been obtained so far \cite{sofar}. Thus, it is tempting to try the present method in more dimensions, exploiting the Lorentz symmetry and the strong subadditive inequality on the geometric entropy. However, one has to bear in mind that no compelling physical principle guarantees that a version of the theorem has to hold in $d\ge 2$ \cite{frad}.  
We have essayed different possibilities without success, since several obstacles show up in higher dimensions. What follows is a list of some of them. They may be avoided in de Sitter space but at the expense of introducing a dimensionfull parameter (the space-time radius) which impedes the proof of the theorem.  

\noindent {\bf a-.} {\sl The shape problem:} a  generalization of the construction on figure 1 requires to select a specific shape for the set replacing  the interval in two dimensions. The first obstacle is then that in $d\ge 2$ the intersections and unions of sets with a given shape are in general of a different shape.  Also, the SSA relation between them in general contains divergent terms on each side which do not cancel due to the new surface features contained in the intersection and the union. This put difficulties in the gain of useful information from the SSA relation.

\noindent {\bf b-.} {\sl The commutativity problem:} there are certain shapes for which the previous problem does not appear, for example two rectangles formed by moving one of them along one of its sides. However, if the rectangles are boosted to each other (this is essential in (\ref{catorce})) there are pieces of the boundary of one of the rectangles which are at a timelike distance from the other: technically they do not commute in the algebra of causal sets, and the SSA relation cannot be used \cite{h1}. We can avoid this problem by using the limit of rectangles with one large side. However, in this case we have to consider intensive quantities with respect to the large side, the corresponding $c$ functions are dimensionfull, and cannot be used to prove the theorem. 

\noindent {\bf c-.} {\sl The order of the divergences:} this is perhaps the deepest problem encountered. Because of its nature, the SSA relation when used between infinitesimally displaced sets, allow us to obtain inequalities involving only second order derivatives of the entropy. However, in $d$ spatial dimensions we have general divergent terms with $d-1$ dimensions. Thus, the local SSA relation cannot give inequalities for universal quantities if $d\ge 2$.

\noindent {\bf d-.} {\sl Logarithmic terms:} the mutual information $I(A,B)$ is universal, dimensionless and increasing with the size of $A$ and $B$. Thus, $I(A,B)$ is decreasing under dilatations for star-shaped non-intersecting sets $A$ and $B$. However $I(A,B)$ diverges logarithmically at critical points due to the logarithmic terms induced by the vertices \cite{ch2}.


\bigskip
 
\begin{thebibliography}{99}
   
 \bibitem{bombelli}
 L.~Bombelli, R.~K.~Koul, J.~H.~Lee and R.~D.~Sorkin,
 Phys.\ Rev.\ D {\bf 34}, 373 (1986);
  G.~'t Hooft,
  Nucl.\ Phys.\ B {\bf 256}, 727 (1985).

\bibitem{ryu}
  S.~Ryu and T.~Takayanagi,
  Phys.\ Rev.\ Lett.\  {\bf 96}, 181602 (2006)
  [arXiv:hep-th/0603001];
  JHEP {\bf 0608}, 045 (2006)
  [arXiv:hep-th/0605073];
  T.~Hirata and T.~Takayanagi,
  arXiv:hep-th/0608213.


\bibitem{phase} T. J. Osborne, M. A. Nielsen, Phys. Rev. A {\bf  66}, 032110 (2002) [arXiv:quant-ph/0202162];
 A. Osterloh, L. Amico, G. Falci, R. Fazio, Nature {\bf 416}, 608 (2002) [arXiv:quant-ph/0202029];
G.~Vidal, J.~I.~Latorre, E.~Rico and A.~Kitaev,
Phys.\ Rev.\ Lett.  {\bf 90}, 227902 (2003)
[arXiv:quant-ph/0211074].


\bibitem{topological}
A.~Kitaev and J.~Preskill 
  [arXiv:hep-th/0510092];
  M.~Levin and Xiao-Gang Wen [arXiv:hep-th/0510613];
  P.~Fendley, M.~P.~A.~Fisher and C.~Nayak,
  [arXiv:cond-mat/0609072].


\bibitem{ch1}
H.~Casini and M.~Huerta,
Phys.\ Lett.\ B {\bf 600}, 142 (2004)
[arXiv:hep-th/0405111].

\bibitem{rg}
See also J.~I.~Latorre, C.~A.~Lutken, E.~Rico and G.~Vidal,
 Phys.\ Rev.\ A {\bf 71}, 034301 (2005)
 [arXiv:quant-ph/0404120];
  J.~Gaite,
  Phys.\ Rev.\ Lett.\  {\bf 81}, 3587 (1998)
  [arXiv:hep-th/9710241]; R. Orus, Phys. Rev. A {\bf 71}, 052327 (2005) [arXiv:quant-ph/0501110].


\bibitem{h1}
  H.~Casini,
  Class.\ Quant.\ Grav.\  {\bf 21}, 2351 (2004)
  [arXiv:hep-th/0312238].

\bibitem{conical} See for example C.~G.~Callan and F.~Wilczek,
 Phys.\ Lett.\ B {\bf 333}, 55 (1994)
[arXiv:hep-th/9401072];
  D.~V.~Fursaev,
  Phys.\ Lett.\ B {\bf 334}, 53 (1994)
  [arXiv:hep-th/9405143].

\bibitem{conformal}
P.~Calabrese and J.~Cardy,
JSTAT {\bf 0406}, P002 (2004)
[arXiv:hep-th/0405152];
  C.~Holzhey, F.~Larsen and F.~Wilczek,
  Nucl.\ Phys.\ B {\bf 424}, 443 (1994)
  [arXiv:hep-th/9403108].

\bibitem{ch2}
  H.~Casini and M.~Huerta,
  arXiv:hep-th/0606256.




 \bibitem{srednicki}
 M.~Srednicki,
 Phys.\ Rev.\ Lett.\  {\bf 71}, 666 (1993)
 [arXiv:hep-th/9303048].


\bibitem{zamo}
A. B. Zamolodchikov, Pis'ma Zh. Eksp. Teor. Fiz. {\bf 43}, 565 (1986);
 JETP Lett. {\bf 43}, 730 (1986).



\bibitem{ch3}
  H.~Casini, C.~D.~Fosco and M.~Huerta,
  J.\ Stat.\ Mech.\  {\bf 0507}, P007 (2005)
  [arXiv:cond-mat/0505563];
  H.~Casini and M.~Huerta,
  J.\ Stat.\ Mech.\  {\bf 0512}, P012 (2005)
  [arXiv:cond-mat/0511014].

\bibitem{sofar}
  J.~L.~Cardy,
  Phys.\ Lett.\ B {\bf 215}, 749 (1988);
  D.~Anselmi,
  Annals Phys.\  {\bf 276}, 361 (1999)
  [arXiv:hep-th/9903059];
  A.~Cappelli and G.~D'Appollonio,
  Phys.\ Lett.\ B {\bf 487}, 87 (2000)
  [arXiv:hep-th/0005115];
  S.~Forte and J.~I.~Latorre,
  Nucl.\ Phys.\ B {\bf 535}, 709 (1998)
  [arXiv:hep-th/9805015];
  A.~Cappelli, R.~Guida and N.~Magnoli,
  Nucl.\ Phys.\ B {\bf 618}, 371 (2001)
  [arXiv:hep-th/0103237].


\bibitem{frad}  
  A.~H.~Castro Neto and E.~H.~Fradkin,
  Nucl.\ Phys.\ B {\bf 400}, 525 (1993)
  [arXiv:cond-mat/9301009].
\end{thebibliography}
\end{document}